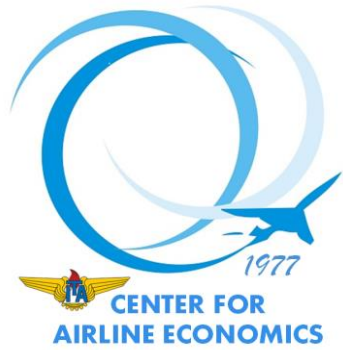

# WORKING PAPER SERIES

Testing the differentiated impact of the COVID-19 pandemic on air travel demand considering social inclusion

Luca J. Santos
Alessandro V. M. Oliveira
Dante Mendes Aldrighi

Institutional Repository / Preprint

Aeronautics Institute of Technology
São José dos Campos, Brazil

# Testing the differentiated impact of the COVID-19 pandemic on air travel demand considering social inclusion

Luca J. Santos

Alessandro V. M. Oliveira✈

Dante Mendes Aldrighi

**Abstract**

The economic downturn and the air travel crisis triggered by the recent coronavirus pandemic pose a substantial threat to the new consumer class of many emerging economies. In Brazil, considerable improvements in social inclusion have fostered the emergence of hundreds of thousands of first-time fliers over the past decades. We apply a two-step regression methodology in which the first step consists of identifying air transport markets characterized by greater social inclusion, using indicators of the local economies' income distribution, credit availability, and access to the Internet. In the second step, we inspect the drivers of the plunge in air travel demand since the pandemic began, differentiating markets by their predicted social inclusion intensity. After controlling for potential endogeneity stemming from the spread of COVID-19 through air travel, our results suggest that short and low-density routes are among the most impacted airline markets and that business-oriented routes are more impacted than leisure ones. Finally, we estimate that a market with 1% higher social inclusion is associated with a 0.153% to 0.166% more pronounced decline in demand during the pandemic. Therefore, markets that have benefited from greater social inclusion in the country may be the most vulnerable to the current crisis.

___________________________________

✈ Corresponding author. Email address: alessandro@ita.br.

▪ Affiliations: Center for Airline Economics, Aeronautics Institute of Technology, Brazil (first two authors); University of São Paulo, Brazil (third author).

▪ Acknowledgments: The second author wishes to thank the São Paulo Research Foundation (FAPESP)–grants n. 2013/14914-4, 2015/19444-1, and 2020-06851; National Council for Scientific and Technological Development (CNPq)–grants n. 301654/2013-1, and 301344/2017-5. The authors wish to thank Rogéria Arantes, Giovanna Ronzani, Cláudio Jorge P. Alves, and Tiago Pereira. All mistakes are ours.

## 1. Introduction

The COVID-19 outbreak triggered a dramatic worldwide shock to air transport demand. The significant health challenges, intensified by the fear of contracting and spreading the Coronavirus,[1] provoked the closure of several international borders and forced an unprecedented wave of flight cancelations. This has strongly impacted commercial aviation and related tourism sectors and brought the risk of a severe economic crisis that may last for many years. In many emerging economies, the short-, medium-, and long-term consequences of the pandemic pose a substantial threat to the middle-income consumer segment, which has expanded considerably over the past few decades.[2]

In Brazil, considerable advances in social inclusion since the 1990s have allowed the emergence of hundreds of thousands of middle class-travelers (Neri, 2015; Klein, Mitchell, and Junge, 2018). Macroeconomic stabilization policies, economic growth, and improvements in living conditions, combined with the industry's economic deregulation, decisively contributed to the greater participation of a rising middle class. Additionally, labor income inequality indicators began to improve in 2019, after having worsened for four years since the start of the recession.[3] However, with the advent of the coronavirus crisis, income inequality has been intensifying across the country, only partially mitigated by the temporary emergency aid grant from the government to more than 60 million people.[4] As a result of the slump in economic activity and the coronavirus-induced health crisis, air travel dropped by 81%—from 45.7 million revenue passengers in the second and third quarters of 2019 to only 8.7 million for the same period in 2020.

This study investigates the impact of the coronavirus crisis on air transportation, focusing on the future dynamics of market positioning in the airline industry. In Brazil, the low-cost carrier (LCC) business model has been losing strength over the years, with only Gol Airlines operating in this segment but without large differentials in ticket prices and mean yield than its primary rival, Latam.[5] Since its merger with the regional trip airlines, Azul Airlines, another incumbent, has been positioning itself more strongly toward high-yield passengers and the monopolistic regional markets. The coronavirus crisis has brought relevant challenges to each of these firms as the traditional passenger segment has changed its purchasing behavior, with business travelers swapping trips for videoconferencing. In this context, in which market segments has the current crisis most impacted non-business passengers' demand? The markets in which non-business passengers strongly participate are usually those related to tourism and the regions marked by greater social inclusion over the past decades.

---

[1] Ornell et al. (2020).

[2] See "An emerging middle class" (Mario Pezzini, 2012) available at www.oecdobserver.org; "The emergence of the middle class: an emerging-country phenomenon" (Clàudia Canals, September 16, 2019) available at www.caixabankresearch.com; "What China's Middle Class Says about Trump, Trade and Tomorrow," February 27, 2019, available at www.bloomberg.com; and "The 'Bird of Gold': The Rise of India's Consumer Market," May 2007, available at www.mckinsey.com.

[3] "After four years, income inequality at work stops worsening" ("Após quatro anos, desigualdade de renda do trabalho para de piorar"), November 20, 2019, available at www.valor.globo.com.

[4] "With the Coronavirus crisis, inequality worsens in metropolitan regions" ("Com crise provocada pelo coronavírus, desigualdade piora nas regiões metropolitanas"), October 22, 2020, available at g1.globo.com; "Brazil faces hard spending choices in 2021," December 19, 2020, available at www.economist.com.

[5] See Wang, Zhang, and Zhang (2018) for the impacts of the presence of LCC on the demand of other emerging airline markets.

Our econometric methodology for identifying markets characterized by greater social inclusion in the Brazilian air transportation industry allows us to quantify the drivers of the pandemic's short-term demand shrinkage throughout 2020, to estimate the demand evolution for the coming years, and to figure out the airlines' attendant business model's needs for adjustment. This methodological framework employs indicators that capture income distribution, credit availability, and internet access at the local level as proxies for social, financial, and digital inclusion factors in the industry, allowing us thereby to estimate their impacts on air travel. We contribute to the literature by carrying out an empirical analysis of the determinants of the sharp drop in demand for air travel since the pandemic began by differentiating the markets according to their estimated intensity of social inclusion. Furthermore, this approach addresses endogeneity issues that arise due to the possible spread of COVID-19 through air travel, a factor that may cause inconsistent estimation and bias in air transport demand coefficients when considering 2020 data. Our results suggest that, after controlling for the intensity of the coronavirus contagion, social inclusion factors are relevant in explaining the fall in traffic on domestic routes over the period.

This paper is organized as follows. Section 2 reviews the relevant literature on the impact of the COVID-19 pandemic on air travel markets; Section 3 describes the Brazilian air transport industry, discusses the improvement in social inclusion indicators and the emergence of first-time air transport consumers, and presents the data sample, the empirical model, and the estimation strategy; Section 4 shows and interprets the estimation results; and Section 5 concludes.

## 2. The impacts of the COVID-19 pandemic on air travel markets

The literature regarding the COVID-19 pandemic's impacts on air transport markets worldwide is thriving. To date, these studies have focused on airlines, air transport-related sectors, and the local economy and seek to discuss government actions to mitigate the associated effects.

Taking the passenger perspective, Monmousseau et al. (2020) investigate the impact of the travel restriction measures implemented during the COVID-19 pandemic on the U.S. airline industry. They compute, close to real time, indicators extracted from social media to measure how travel restrictions impacted the relationship between passengers and airlines. Iacus et al. (2020) develop a forecasting model to project air passenger traffic during the pandemic. They document that in the first quarter of 2020, the impact of aviation losses may have translated to a global GDP decline of 0.02% to 0.12%. Maneenop and Kotcharin (2020) estimate the short-term impact of COVID-19 on the stock returns of 52 listed airline companies worldwide using an event study methodology. Focusing on the effects of historical uncertainty shocks on airline employment, Sobieralski (2020) finds that job loss is nearly 7% in the airline workforce, with an upper bound of over 13%, and that recovery following the shocks takes between four and six years. Brown and Kline (2020) study the managerial preparedness of U.S. airlines in the COVID-19 pandemic and conclude that airline management teams failed to learn from previous outbreaks. Czerny et al. (2020) focus on the post-pandemic recovery of the Chinese aviation market.

Macilree and Duval (2020) discuss International Civil Aviation Organization ICAO's role in coordinating safety provisions during the pandemic and nationwide state aid in airline bailouts, recapitalization, and ownership. Abate et al. (2020) study government support for airlines in the aftermath of the COVID-19 pandemic, indicating that governments prioritize sustaining air transport connectivity to protect economic activity and jobs in both the aviation and tourism sectors.

Naboush and Alnimer (2020) investigate the circumstances under which an air carrier is liable for the transmission of COVID-19 and the scope of the safety measures required by the ICAO to prevent its spread. Prince and Simon (2020) focus on the geographic heterogeneity of the impact of international travel on the spread of COVID-19 in the United States. Similarly, Nakamura and Managi (2020) calculate the overall relative risk of COVID-19 importation and exportation for every airport in local municipalities worldwide.

By contrast, Lamb et al. (2020) consider travelers' perspectives and investigate the factors associated with their willingness to fly during the COVID-19 pandemic. They survey 632 participants in the United States, extracting characteristics of demographics, personality, emotional state, and travel purposes, and they develop regression models for both business and pleasure travel. They find that the perceived threat from COVID-19, agreeableness, affect, and fear are statistically significant regressors for both the business and pleasure types of fliers. The absolute value of their estimated standardized coefficients associated with "the perceived threat from COVID-19" and "fear" variables are higher for pleasure travelers than for business travelers: -0.172 against -0.159 for "perceived threat from COVID-19" and -0.063 against -0.050 for "fear." Their approach allows us to better understanding the passenger behavior in the aftermath of the crisis. With the authors' breakdown of passenger types—business and pleasure segments—it is possible to project adjustments to market positioning and even restructure the airline business models in the aftermath of the crisis.

In line with Lamb et al. (2020), we perform a market segmentation for analysis. However, unlike them, we focus on an emerging market country's air transport industry, wherein social inclusion is a critical factor in determining the consumer segments' relative participation: the higher the social inclusion, the lower the participation of the mainstream segment of business-related travelers. In contrast to their study, our econometric approach uses panel data from city pairs across the country to identify the routes that rely most on middle-class consumers and to investigate the vulnerability of these market segments to the impacts of the economic crisis brought about by the pandemic. Based on Prince and Simon (2020), who suggest that geographic heterogeneity of the spread of COVID-19 in the U.S. may be related to population density, cultural differences, public response, and potential virus carriers' inflow, we use a geographic heterogeneity lens to understand the effects of the pandemic on air transport demand.

The surveyed literature guides our empirical approach design. Based on the findings of Monmousseau et al. (2020), insights can be obtained regarding the demand induction factors resulting from greater digital inclusion that were possibly relevant to mitigating the fall in demand observed during the pandemic period. The digital inclusion factor is one of the key aspects considered in our work. The demand forecast of Iacus et al. (2020) motivated us to extend the literature to decompose the observed drop in air passenger traffic into its possible determinants. From the study by Maneenop and Kotcharin (2020) and Brown and Kline (2020), we can see the market pressures on airlines from the capital markets and management perspectives, respectively. Under such circumstances, airlines certainly need to adjust their business models to suit the new post-pandemic market reality—implications that we consider when interpreting our estimation results. With the analyses of Macilree and Duval (2020) and Christidis and Purwanto (2020), it is possible to see the impacts of public institutions' interventions on air transport during the crisis. We also consider variables related to public and corporate policies, such as the impact of the Brazilian government's emergency financial aid grant and the "Essential Air Network," conceived by the airlines and the government. Finally, inspired by Naboush and Alnimer (2020), Nakamura and Managi (2020), and Lamb et al. (2020), we study the role of the number of COVID-19 cases and its possible endogeneity with air transport demand.

## 2.1. Social inclusion, demand for air travel, and the impacts of the COVID-19 pandemic

Traditionally, air travel demand is driven by factors related to income and price elasticities of demand (Gillen, Morrison, and Stewart, 2007). In general, when economic activity accelerates, passenger travel is induced for business reasons and, as the disposable income of families increases, for their higher spending on leisure travel. With increased competition among airlines, prices fall, and thus the number of trips rises. However, in many emerging countries, poor national income distribution can hinder the air transport sector's expansion. As sustainable economic development is promoted with less concentrated income, a potentially significant emerging middle class may gain access to the air travel market, promoting traffic growth. This phenomenon is certainly in the context of the advancement of social inclusion in developing countries. According to the World Bank, "Social inclusion is the process of improving the terms on which individuals and groups take part in society—improving the ability, opportunity, and dignity of those disadvantaged on the basis of their identity."[6]

The United Nations emphasizes the role of some forms of social progress and social inclusion, highlighting issues of financial inclusion (access to useful and affordable financial products and services)[7] and digital inclusion (access and use of information and communications technologies).[8] According to the United Nations Department of Economic and Social Affairs, these issues have become even more relevant with "successive lockdowns and measures confinement put in place by governments to curb the spread of COVID-19."[9]

Before the pandemic, with the reality of continuous social progress, the growth of the middle class played a crucial role in driving structural changes in the economies' consumption patterns as people came out of poverty. According to a consultancy report, from 1990 to 2005, the size of the middle class impressively grew from 1.4 to 2.6 billion worldwide.[10] The rise of the middle class in Asia, Latin America, and Africa has boosted the propensity to fly in these regions. For this reason, the International Air Transport Association estimated in 2015 deliveries of more than 6,000 new jets with a capacity of between 70 and 130 seats by 2035 in the region.[11]

With the advent of health and economic crises triggered by the COVID-19 pandemic, much uncertainty has emerged regarding the continuation of the spillover effects of social inclusion on air transport demand that had marked many emerging markets for years. The abrupt drop in air travel in 2020, although global, has had varying intensities across countries and has been dictated by the unpredictable advancement of new coronavirus variants. Additionally, government restrictions as policy responses, such as business closures and lockdowns, have been heterogeneous and, in many cases, far from efficient. Ferraresi et al. (2020) find that countries characterized by low development and digitalization levels, among other factors, have adopted less stringent measures. As a result, the higher levels of unemployment and impoverishment brought about by the continuation of the pandemic status quo may produce dramatic and enduring socioeconomic consequences in many countries. In this context, a plausible hypothesis is that the pandemic may have so far been more harmful for the air travel demand in markets previously characterized by recent advances in social inclusion. Our study seeks to examine this hypothesis empirically in the context of domestic air transport in Brazil.

---

[6] Source: www.worldbank.org/en/topic/social-inclusion.

[7] Source: www.worldbank.org/en/topic/financialinclusion.

[8] Source: http://mediawiremobile.com/news/what-is-digital-inclusion-and-why-is-it-important.

[9] "Leveraging digital technologies for social inclusion," The United Nations Department of Economic and Social Affairs, available at https://www.un.org/development/desa/dspd/wp-content/uploads/sites/22/2021/02/PB_92-1.pdf.

[10] This is from Ernst Young (2013) "Hitting the sweet spot - The growth of the middle class in emerging markets," available at iems.skolkovo.ru.

[11] "Rise of the middle class," SP's aviation, June 2015, available at www.sps-aviation.com.

**3. Research design**

Our methodology for identifying air transport markets related to greater social inclusion relies on data for the period when social inclusion significantly advanced in Brazil. Since the macroeconomic stabilization in the 1990s, Brazil has experienced spurts of economic growth and welfare improvement for the less favored classes. At the same time, income transfer programs have helped combat extreme poverty in the country, while a declining income inequality gave rise to the emergence of a "new middle class" (Neri, 2015), who started consuming new goods as well as basic health, education, financial, and technology services. Combined with the airline industry's deregulation in the early 2000s, this social development hugely contributed to the emergence of hundreds of thousands of new air transport consumers in the 1990s and the 2000s. Air transport has become less associated with consumer elites, becoming a more frequent item in the consumption basket of many households. Nonetheless, the economic downturn in the mid-2010s, along with the coronavirus outbreak in 2020, has interrupted such trend.

*3.1. The coronavirus outbreak in Brazil*

On March 11, 2020, the World Health Organization (WHO) characterized the spread of the disease caused by the new Coronavirus, COVID-19, as a pandemic.[12] Officially, the first cases of the disease appeared in China in December 2019, but soon after, new cases were diagnosed in Europe and other parts of Asia, subsequently spreading to the rest of the world. In Brazil, the first official record of the disease was on February 26, 2020.[13] On March 17, 2021, COVID-19 had already caused more than 284,000 deaths in the country.[14]

Commercial aviation's total revenue passenger kilometers (RPK) grew in Brazil by 0.6% in 2019.[15] The record passenger movement in January 2020 (more than 9.8 billion RPKs), as compared with

---

[12] "WHO characterizes COVID-19 as a pandemic," March, 11 2020, available on www.who.int.
[13] "Coronavirus timeline in Brazil" ("Linha do tempo do coronavírus no Brasil"), February 26, 2020, available on Coronavirus.saude.gov.br.
[14] "Coronavirus panel" ("Painel Coronavírus"), accessed on January 7, 2021, available on COVID.saude.gov.br.
[15] "Air transport demand and supply - ASK, RPK and Utilization - Reference: last 12 months of December 2019."

January in the previous years,[16] raised optimistic performance expectations for the sector. However, the alarming diffusion of the new Coronavirus in Brazil in early 2020 led the Ministry of Health to launch precautionary measures to deter its spread over the country, which soon proved insufficient. This sequence of events forced the government to announce mandatory quarantine in mid-March.[17]

The severity and high rate of contagion of the disease and the government's attendant sanitary measures to contain it dramatically affected most sectors of the economy and cast uncertainty in the near future. The tourism-related sectors have been among the most affected by the pandemic, notably air travel, due to the perception that they could be a conduit to the virus's spread. As in other countries, Brazilian airlines had to deal with an unprecedented adverse demand shock. Passenger traffic fell sharply with the nationwide spread of the disease. According to the National Civil Aviation Agency (ANAC),[18] Brazilian airlines suffered a record 94.5% plunge in passenger movement in April 2020 as compared with the same period in 2019. The aircraft occupancy rate also registered a sharp drop in the same period, from 81.9% to 65.4%.[19] Moreover, losses of leading airlines exceeded R$ 15 billion in the first six months of 2020.[20]

In March and April 2020, the government and airlines tried to tackle the crisis by relaxing slots rules at controlled airports (i.e., slots waiver), postponing airport concessionaires' contractual payments for the concession, and implementing an "essential airline network," among other measures.[21] In search of solutions to accelerate recovery and avoid further losses, Azul and Latam signed a codeshare agreement in August.[22] Combining Latam's presence in the country's leading hubs and Azul's capillarity on flights to destinations with lower demand and infrastructure support, the agreement expanded their network, and the frequency of flights offered to passengers.

*3.2. Sources of data*

We rely on two datasets. The first dataset is a panel of directional city pairs of the domestic airline industry in Brazil, with monthly observations between July 2010 and December 2018. We include only passenger flights. To compute a city pair, we group multiple airports in the same catchment area. We only consider routes with six or more observations and a hundred passengers in each period. The full sample size is 65,452, but it reduces to 48,957 in some specifications owing to the social inclusion regressor's data availability. The second dataset reports only data for 2020 and comprises 763 original city pairs that still had scheduled flights in 2019. To examine thoroughly the drivers of the notable decline in air travel demand in 2020, we consider only the first full quarters since the pandemic outbreak. We then develop a cross-section of routes considering the second (Q2) and third (Q3) quarters of 2020 and compare the figures of 2020Q2–Q3 with 2019Q2–Q3 to analyze the pandemics' short-run effects on the air travel markets.

---

[16] "Air transport demand and supply - ASK, RPK and Utilization - Reference: January 2020."

[17] Diário Oficial da União, Portaria Nº 356, 11 de março de 2020, available on www.in.gov.br.

[18] "iNFRADebate: Impacts of the pandemic on civil aviation - What are the horizons for the resumption of the sector" ("iNFRADebate: Impactos da pandemia na aviação civil - Quais os horizontes para a retomada do setor?") by Fabiana Todesco e Mauro Césa Santiago Chaves, September 3, 2020, available on www.agenciainfra.com.

[19] "Air transport demand and supply - ASK, RPK and Utilization - Reference: April 2019 and 2020," with the author's own calculations.

[20] "2nd quarter data reveal impact of more than 6 billion on Brazilian airlines" ("Dados do 2º tri revelam impacto de mais de 6 bilhões nas aéreas brasileiras"), October 23, 2020, available on www.anac.gov.br.

[21] "Main measures in the airline industry after the beginning of the COVID-19 pandemic - Timeline" ("Principais medidas do setor aéreo após início da pandemia de COVID-19 - Linha do Tempo"), June 2, 2020, available on www.anac.gov.br.

[22] "Azul and Latam start codeshare: understand how it works" ("Azul e Latam iniciam codeshare: entenda como funciona"), G1, August 12, 2020, available on g1.globo.com.

Most data for both datasets are available from the ANAC in online public databases.[23] The agency provides information on all domestic and international scheduled flights on its air transportation market statistics database and its active scheduled flight report (VRA). For the social inclusion proxies, we use the following sources. First, regarding the standards of living, health, and education, we use data from Firjan - the Industry Federation of the State of Rio de Janeiro -, specifically, the Firjan Municipal Development Index.[24] For the financial inclusion proxies, we use the Central Bank of Brazil's ESTBAN database, a publicly available dataset on the number of branches, and a detailed balance sheet for each local bank at the city-month level. We use data from the National Telecommunication Agency (ANATEL) for the digital inclusion proxies at the city-month level. Other data sources include the Brazilian Institute of Geography and Statistics (IBGE), the National Agency for Petroleum, Natural Gas and Biofuels (ANP), the National Land Transport Agency (ANTT), the Ministry of Health (regarding coronavirus cases), and the Brazilian government's Transparency Portal[25] on emergency aid benefits.

### 3.3. *Empirical model*

Equations (1) and (2) present the specifications of our proposed empirical model. First, Equation (1) presents our model of air travel demand in Brazil (first data set):

$$\begin{aligned} \text{PAX}_{k,t} = &\ \beta_1 \text{INC}_{k,t} + \beta_2 \text{P}_{k,t} + \beta_3 \text{PBUS}_{k,t} + \beta_4 \text{TOUR}_{k,t} + \beta_5 \text{NET}_{k,t} + \\ &\ \beta_6 \text{P}_{k,t} \times \text{LEIS}_{k,t} + \beta_7 \text{TREND}_t + \beta_8 \text{TREND}_t \times \text{REC}_t + \\ &\ \beta_9 \text{HDI(DSL)}_{k,t} + \beta_{10} \text{CELL}_{k,t} + \beta_{11} \text{LOAN}_{k,t} + \beta_{12} \text{DEBT}_{k,t} + u_{k,t}, \end{aligned} \qquad (1)$$

---

[23] See www.nectar.ita.br/avstats for a description and access to all air transport datasets in Brazil.
[24] See Avelino, Bressan, and Cunha (2013) for a description of the Firjan index.
[25] www.portaltransparencia.gov.br.

and second, Equation (2) presents a framework to pinpoint the effects of the COVID-19 pandemic on air travel demand (second data set):

$$\begin{aligned} \text{PAXVAR}_k = \sum_j \gamma_j \text{DIST}_k^{(j)} &+ \delta_1 \text{PVAR}_k + \delta_2 \text{INCPRE}_k + \delta_3 \text{DENSPRE}_k + \\ &\delta_4 \text{LEISPRE}_k + \delta_5 \text{ESSENTNET}_k + \delta_6 \text{ESSENTNET} \times \text{AMAZ}_k + \\ &\delta_7 \text{CODESHARE}_k + \delta_8 \text{COVID}_k + \delta_9 \text{COVID}_k \times \text{DIST}_k + \\ &\delta_{10} \text{COVID}_k \times \text{EMERGAID}_k + \delta_{11} \text{COVID}_k \times \text{SOCINCL}_k + \varepsilon_k, \end{aligned} \qquad (2)$$

where $k$ denotes the city-pair, $t$ the periods ($t = 1, ..., 102$ months). $u_{k,t}$ and $\varepsilon_k$ are the error terms, and the $\beta$'s, $\gamma$'s and $\delta$'s are the parameters to be estimated. In Table 1, we provide a presentation of the variables in Equations (1) and (2). Detailed information on each variable is available in the Appendix.

### *3.4. Estimation strategy*

The empirical approach of the present study involves estimating the two demand equations. Equation (1), which has PAX as the regressand, estimates the effects of social, digital, and financial inclusion factors that were possibly very relevant in generating air travel demand in the early 2010s. For this purpose, we employ the regressors HDI (DSL), CELL, LOAN, and DEBT. In the first specification, we use HDI—the overall Human Development Index of the endpoint cities—to capture the effects of socioeconomic factors in the most general way. Equation (2), which has PAXVAR as the regressand, has the form of demand variation, to capture the main drivers of the remarkable fall in demand for air travel in the first two quarters of the pandemic (2020Q2–Q3) as compared with the same period of the previous year. For this equation, we focus on the results of the regressor COVID and its interactions (COVID × DIST, COVID × EMERGAID, and COVID × SOCINCL). Through interaction variables, it is possible to test the existence of moderation effects (intensifiers or attenuators) on the relationship between COVID and PAXVAR in the sample period. We focus on the hypothesis test of the estimate of COVID × SOCINCL because SOCINCL is calculated from a prediction using the estimated coefficients of variables HDI (DSL), CELL, LOAN, and DEBT in Equation (1). The idea of this procedure is to test whether the routes assigned with the most prominent social, digital, and financial inclusion indicators in the early 2010s would have some differentiated impact caused by the COVID-19 pandemic.

Table 1 - Description of model variables[26]

| Variable | Description |
|---|---|
| **Equation (1)** | |
| PAX | total airline tickets sold (ln) |
| INC | per capita gross domestic product (ln) |
| P | mean airline price (ln) |
| PBUS | mean bus price (ln) |
| TOUR | a proxy for the size of the tourism market (ln) |
| NET | number of cities served (ln) |
| LEIS | high season period (dummy) |
| TREND | time trend |
| REC | mid-2010s recession (dummy) |
| HDI(DSL) | Human Development Index (HDI) 's dimension for "decent standard of living", as a proxy for social inclusion (ln) |
| CELL | cell phones, as a proxy for digital inclusion (ln) |
| LOAN | households' access to credit, as a proxy for financial inclusion (ln) |
| DEBT | household indebtedness, as a proxy for financial inclusion (ln) |
| **Equation (2)** | |
| PAXVAR | variation in total air tickets sold between the two first quarters after the pandemic, 2020, and the same period in the previous year, 2019 (fraction) |
| DIST$^{(j)}$ | j-th route distance interval: 0 - 500 km (base case), 500 - 1000 km, 1000 - 1500 km, 1500 - 2000 km, 2000 km or higher (dummies) |
| PVAR | variation in mean air travel price between the two first quarters since the pandemic outbreak, 2020, and the same period in the previous year, 2019 (fraction) |
| INCPRE | per capita gross domestic product previous to the pandemic |
| DENSPRE | route density previous to the pandemic |
| LEISPRE | proportion of leisure passengers previous to the pandemic |
| ESSENTNET | "Essential Air Network", an emergency play carried out by the government and the airlines |
| AMAZ | Amazon state countryside (dummy) |
| CODESHARE | Latam-Azul airlines codeshare agreement operations (dummy) |
| COVID | confirmed COVID-19 infection cases (ln) |
| DIST | route distance in kilometers (ln) |
| EMERGAID | emergency financial aid grants (ln) |
| SOCINCL | a proxy for social inclusion, predicted from Equation (1) 's coefficients of HDI(DSL), CELL, LOAN, and DEBT. |

---

[26] Note that we omit subscripts *k* and *t*. See details of each variable in the Appendix.

Our two-step procedure can be summarized as follows. In the first step, we estimate the social inclusion parameters of demand in Equation (1). In the second step, we calculate the estimated social inclusion metric SOCINCL and plug it into Equation (2) as an interaction with COVID. The estimation problem here is that the standard errors of the second-step parameters must be adjusted to reflect that the social inclusion variables are previously estimated. To accomplish this, we compute bootstrap-adjusted standard errors computed using a stratified bootstrapping procedure with 2,000 replications.

Our empirical approach addresses two additional specification challenges. First, we need to account for unobserved factors that drive air travel demand across a large country like Brazil. Undoubtedly, many regional idiosyncrasies either stimulate or hinder demand that our framework of a few covariates in Equations (1) and (2) is unable to control for. For example, in Equation (2), we suspect that psychological effects related to the fear of the coronavirus contagion may drive the assessment of risk and the purchase behavior of consumers (Lamb et al., 2020; Kim et al., 2020), and, for a theoretical framework, we refer to Kahneman and Tversky (1979). Additionally, although we control for the government's emergency financial aid granted since the first months of the pandemic, the actual evolution of local economies is a critical unobserved factor of our empirical model. Therefore, we consider an amplified set of control variables in our robustness checks to account for such unobservables in our analysis. In Equation (1), we employ region-specific seasonal dummies, local temperature, and panel time dummies to account for local seasonality and global time effects. In Equation (2), we use regional and region-interacted dummies to account for specificities that are particular to the endpoint airports' regions of the routes. We provide details of these settings in the robustness check section.

The second specification challenge is endogeneity. In Equations (1) and (2), we have a typical problem of demand-side endogeneity of the right-hand side price variables: P, P × LEIS, and PVAR. Therefore, we need proper instrumental variables (IV) to proceed with a consistent estimation. For these variables, we use cost shifters, Hausman-type, Stern-type, and BLP-type IVs (See Mumbower, Garrow, and Higgins (2014) for a detailed discussion of these classes of IVs). All IVs are computed in natural logarithms.

A more subtle endogeneity issue is related to the pandemic's effect on the decline in air travel intensity. In Equation (2), COVID is potentially correlated with the unobserved factors that drive PAXVAR. The endogeneity of COVID is due to the possible spread of the SARS-CoV-2 virus through air travel (Adiga et al., 2020; Barnett and Fleming, 2020; Nakamura and Managi, 2020). Schuchat et al. (2020) describe the continued travel-associated importation of the Coronavirus from outside the U.S. as an early contributor to the initiation and acceleration of domestic COVID-19 cases. Prince and Simon (2021) investigate the effect of international air travel on the spread of COVID-19 in the United States. They consider environmental and exposure factors that explain the substantial geographic heterogeneity of the COVID-19 spread. They employ population density, cultural differences, and public response proxies as environmental factors, and the inflow of international airline travelers as exposure factors. They find statistically significant impacts of the inflow of international airline travelers on the COVID-19 spread in the U.S. territory when the air traffic originated from pandemic hotspots.

As the coronavirus contagion is possibly positively correlated with the unobserved dynamics of the local economies, we consider that COVID may be correlated with the error term $\varepsilon_{kt}$ of Equation (2), thus raising the risk of inconsistent estimation—more specifically, a positive bias that may underestimate the downward effect of the COVID variable. To address this issue, we utilize other factors that may be correlated with the spread of COVID-19, such as the endpoint cities' population densities, areas, and public health and education conditions.[27] We also use these variables to interact with the LEIS as additional instruments. Finally, we employ the origin and destination's minimum, mean, median, and maximum local temperatures as instruments.[28] We build upon our IV strategy on recent studies suggesting some of the factors that might contribute to the spread of Coronavirus, either directly or indirectly. Schuchat et al. (2020) point to high population density as a probable contributor to the spread, among other factors. Pramanik et al. (2020) and Tosepu et al. (2020) find evidence of the role of climatic predictors. Kejela (2020) suggests factors such as inadequate knowledge and public hygiene conditions, the society's socioeconomic nature, and the knowledge, attitude, and practice of health care workers. Robertson (2021) employs proxies for pre-existing health conditions in the population and local economic and demographic conditions, among others.

## 4. Estimation results

Tables 2 and 3 present the results of our empirical models of air travel demand (PAX) and air travel demand decline due to the COVID-19 pandemic (PAXVAR) in the Brazilian airline industry. To simplify the exposition of the results, we omit indexes *k* and *t* of the model variables.

---

[27] Source: IBGE and Firjan, with the author's calculations.

[28] We also create gravity versions of these variables, namely, the minimum, maximum, and geometric mean between the origin and destination of each route at each period (Source: Meteorological Database for Education and Research of the Brazilian National Institute of Meteorology; BDMEP/INMET). We thank an anonymous reviewer for this suggestion.

**Table 2 - Estimation results: air travel demand (PAX)**

|  | (1) PAX | (2) PAX | (3) PAX | (4) PAX | (5) PAX | (6) PAX | (7) PAX | (8) PAX | (9) PAX |
|---|---|---|---|---|---|---|---|---|---|
| INC | 1.8229*** | 1.8614*** | 1.4219*** | 1.4621*** | 1.2579*** | 1.3145*** | 1.2267*** | 0.9260*** | 0.4603*** |
| P | -1.3307*** | -1.3069*** | -1.3200*** | -1.3381*** | -1.3020*** | -1.3118*** | -1.3616*** | -1.4593*** | -1.5521*** |
| PBUS | 0.4355*** | 0.2625*** | 0.5161*** | 0.5267*** | 0.6626*** | 0.4388*** | 0.4984*** | 0.1606** | 0.1518* |
| TOUR | 0.0059*** | 0.0064*** | 0.0070*** | 0.0071*** | 0.0086*** | 0.0083*** | 0.0077*** | 0.0057*** | 0.0055*** |
| NET | 0.1751*** | 0.1263*** | 0.1010*** | 0.1066*** | 0.1311*** | 0.1412*** | 0.1207*** | 0.1072*** | 0.1295*** |
| P × LEIS |  |  |  |  | -0.0117*** | -0.0234*** | -0.0236** | -0.0515*** | -0.0387*** |
|  |  |  |  |  |  |  |  |  |  |
| TREND | -0.0044*** | -0.0042*** | -0.0042*** | -0.0030*** | -0.0022*** | -0.0022*** | -0.0026*** | -0.0029*** |  |
| TREND × REC |  |  |  | -0.0012*** | -0.0017*** | -0.0017*** | -0.0016*** | -0.0012*** |  |
|  |  |  |  |  |  |  |  |  |  |
| HDI |  | 1.3169*** |  |  |  |  |  |  |  |
| HDI (DSL) |  |  | 0.5366*** | 0.3833*** | 0.4091*** | 0.3708*** | 0.4935*** | 0.6736*** | 0.2996*** |
| CELL |  |  | 0.6071*** | 0.5076*** | 0.5313*** | 0.4159*** | 0.5463*** | 0.6595*** | 0.7381*** |
| LOAN |  |  | -0.0001 | -0.0007 | -0.0010* | 0.0027*** | -0.0012** | -0.0010* | -0.0005 |
| DEBT |  |  | -0.0460*** | -0.0455*** | -0.0296*** | -0.0285*** | -0.0261** | -0.0196* | -0.0749*** |
|  |  |  |  |  |  |  |  |  |  |
| Estimator | FE/IV/LASSO | FE/IV/LASSO | FE/IV/LASSO | FE/IV/LASSO | FE/IV/LASSO | FE/IV/LASSO | FE/IV/LASSO | FE/IV/LASSO | FE/IV/LASSO |
| Alternative Version | No | No | No | No | No | SUBSET | No | No | No |
| City-Pair Clusters | 1,041 | 1,013 | 1,013 | 1,013 | 1,013 | 1,013 | 1,013 | 1,013 | 1,013 |
| Local Temp Controls | No | No | No | No | No | No | 15/48 | 11/48 | 11/48 |
| Reg/Seas Controls | No | No | No | No | No | No | No | 250/300 | 242/300 |
| Panel Time Controls | No | No | No | No | No | No | No | No | 72/102 |
| IV Count | 20/438 | 16/438 | 16/438 | 16/438 | 43/876 | 26/876 | 43/876 | 42/876 | 41/876 |
| AIC Statistic | 19,572 | 6,611 | 5,530 | 5,847 | 4,326 | 5,449 | 4,596 | 3,586 | 3,649 |
| BIC Statistic | 19,626 | 6,673 | 5,618 | 5,944 | 4,431 | 5,555 | 4,834 | 5,987 | 6,586 |
| Adj R2 Statistic | 0.4800 | 0.4693 | 0.4809 | 0.4776 | 0.4936 | 0.4818 | 0.4893 | 0.5022 | 0.5022 |
| RMSE Statistic | 0.2832 | 0.2615 | 0.2586 | 0.2595 | 0.2555 | 0.2584 | 0.2562 | 0.2522 | 0.2521 |
| RMSE CV Statistic | 1.1862 | 1.2378 | 1.2703 | 1.2574 | 1.2376 | 1.289 | 1.2533 | 1.3082 | 1.3218 |
| Nr Observations | 65,452 | 48,957 | 48,957 | 48,957 | 48,957 | 48,957 | 48,748 | 48,748 | 48,748 |

*Notes: Estimation results produced by the instrumental variables, post-double-selection LASSO-based methodology of Belloni et al. (2012, 2014a,b), with fixed effects (IV-LASSO). Post-LASSO estimation is performed with a Two-Stage Least Squares, fixed-effects, procedure with standard errors robust to heteroskedasticity and autocorrelation. LASSO penalty loadings account for the clustering of city pairs. Control variables' estimates omitted. Blank cells indicate that the variable was not used. INC, P, PBUS, TOUR, and NET are not penalized by LASSO. Endogenous variables: P and P × LEIS. In Column (6), SUBSET means that the model employs alternative versions of the following subset of variables: PBUS, TOUR, and NET. Cross-validation was performed with a 4-fold procedure. P-value representations: \*\*\*p<0.01, \*\* p<0.05, \* p<0.10.*

Table 2 shows the estimation results for the demand for air travel using data from 2010 to 2018. The nine columns of the table show the coefficient estimation results from using the FE-IV-LASSO procedure. Column (1) presents the baseline model, which contains no social inclusion proxies. This column shows that the estimated income (INC) and price (P) elasticities of demand are 1.8229 and -1.3307, respectively, indicating a relatively high sensitivity of demand to these variables. The PBUS, TOUR, and NET variables have positive, statistically significant estimated coefficients of 0.4355, 0.0059, and 0.1751, respectively, indicating that increases in these factors are associated with demand increases. The trend variable TREND suggests a persistent decrease in demand over time, with an estimated coefficient of -0.0044. These results remain relatively invariant throughout the remaining columns in Table 2.

In Column (2) of Table 2, we insert our first social inclusion metric—the Human Development Index regressor (HDI). The estimated coefficient of this variable is positive and statistically significant, suggesting that enhancements in quality of life and socioeconomic conditions are associated with a higher market size for air transport. In Column (3), we present a breakdown of social inclusion indicators to capture phenomena related to improvements over the geographic space and time in metrics of life conditions: HDI (DSL), digital inclusion (CELL), and financial inclusion (LOAN and DEBT variables). The results in Column (3) and the remaining columns suggest that some of these factors are statistically related to the air travel demand in the sample. The variables HDI (DSL), CELL, and DEBT have statistically significant estimated coefficients; the first two are positive and the last negative. These results indicate that demand is apparently positively driven by socioeconomic factors associated with social inclusion, such as income distribution and access to the Internet: HDI (DSL) and CELL. The result related to household indebtedness (DEBT) shows that financial inclusion has the downside of increasing household indebtedness and discouraging air travel. It is important to note that the LOAN variable, another proxy for financial inclusion through access to bank credit, is not statistically significant in most specifications; this provides some evidence that families did not count on bank credit for airline ticket purchases over the sample period.

In Columns (4) and (5) of Table 2, we insert interaction variables to capture possibly more complex phenomena associated with the demand for air travel: a break in the trend due to the 2014–2016 recession (TREND × REC) and a possible moderating effect of price elasticity of demand on routes with a higher proportion of leisure travelers (variable P × LEIS). Both estimated coefficients are negative and statistically significant, suggesting an intensification role for the respective variables. We consider the model in Column (5) as our preferred specification.

Table 3 presents the estimation results of Equation (2), which is representative of the remarkable drop in demand for air travel due to the COVID-19 outbreak (PAXVAR). Again, we present the results in nine columns, with the first specification being the baseline model. In all versions, we employ the IV-LASSO estimator. As discussed in the empirical strategy section, we apply stratified bootstrapping to adjust the standard error of estimates in the specifications in which we include the social inclusion variables (SOCINCL).[29]

The estimation results in Column (1) of Table 3 indicate the positive effects of distance with respect to PAXVAR. In fact, the estimated coefficients of variables DIST 500–1000, DIST 1000–1500, DIST 1500–2000, and DIST 2000 HIGH are all positive and statistically significant, allowing us to infer that longer routes suffered less demand fall than shorter routes. Additionally, as expected, PVAR has a negative estimated coefficient (-0.5759), suggesting that airlines could partially manage the intensity of the demand drop with price adjustments. Regarding the variables of the routes' characteristics before the pandemic, namely, INCPRE, DENSPRE, and LEISPRE, we observe that only the last two are statistically significant and positively related to PAXVAR. Concerning the essential air network outlined by the major carriers and the government, we have some evidence that only the most inaccessible region of the Brazilian Amazon has benefited from the commitment to the maintenance of operations, as indicated by the results of the variables ESSENTNET (nonsignificant) and ESSENTNET $\times$ AMAZ (positive and statistically significant). Finally, the results in Column (1) point to an apparent absence of statistical significance of the COVID variable: the estimated coefficient of this variable is negative but statistically nonsignificant in Column (1).

In Column (2) of Table 3 we insert the interaction variable COVID $\times$ DIST. Comparing the results of Columns (1) and (2) reveals some interesting changes. First, the estimated coefficient of COVID maintains its negative sign, but becomes statistically significant at the 1% level. This result is clearly associated with the introduction of COVID $\times$ DIST, which has a statistically significant estimated coefficient with a positive sign (0.0328). These results indicate that the downward effect of the spread of COVID-19 on the air transport demand is attenuated by route distance: the longer the route, the lower the decline in demand caused by the pandemic. This finding is consistent with consumers' difficulty in substituting air travel with road trips in medium-to-long-haul markets.[30] It is thus in line with Bauer et al. (2020), who provide evidence that airline business models based on long-haul operations may gain competitive advantage in the COVID context and in the adjustments to the new-normal.

---

[29] Note that SOCINCL was predicted in a previous step, using the model presented in Table 1, Column (4).

[30] Consistent with this result, see "Air travel takes back seat to road trips in the summer of Coronavirus: Fuel for Thought," S&P Global, May 20, 2020, available on www.spglobal.com.

# Table 3 - Estimation results: air travel demand change due to the COVID-19 pandemic (PAXVAR)

|  | (1) PAXVAR | (2) PAXVAR | (3) PAXVAR | (4) PAXVAR | (5) PAXVAR | (6) PAXVAR | (7) PAXVAR | (8) PAXVAR | (9) PAXVAR |
|---|---|---|---|---|---|---|---|---|---|
| DIST 500 - 1000 | 0.0606*** | -0.1932*** | -0.0872*** | -0.0938*** | -0.0942*** | -0.1573*** | -0.1490*** | -0.1648*** | -0.1686*** |
| DIST 1000 - 1500 | 0.1189*** | -0.3091*** | -0.1285** | -0.1407*** | -0.1349** | -0.2371*** | -0.2229** | -0.2477*** | -0.2513*** |
| DIST 1500 - 2000 | 0.1600*** | -0.3796*** | -0.1548** | -0.1731** | -0.1566** | -0.3000*** | -0.2697** | -0.3216*** | -0.3165*** |
| DIST 2000 HIGH | 0.1971*** | -0.4834*** | -0.2046** | -0.2295*** | -0.2169*** | -0.3757*** | -0.3581*** | -0.4078*** | -0.4018*** |
| PVAR | -0.5759*** | -0.2621*** | -0.4827*** | -0.4796*** | -0.4746*** | -0.6835*** | -0.6014** | -0.6956*** | -0.7851*** |
| INCPRE | -0.1133 | 0.0290 | 0.0051 | -0.0142 | 0.0034 | 0.0009 | 0.0659 | 0.0309 | -0.0057 |
| DENSPRE | 0.0359*** | 0.0272*** | 0.0309*** | 0.0245*** | 0.0278*** | 0.0275*** | 0.0326*** | 0.0260*** | 0.0370*** |
| LEISPRE | 0.1523*** | 0.1776*** | 0.1205*** | 0.1226*** | 0.1211*** | 0.0947* | 0.0954* | 0.0822 | 0.0658 |
| ESSENTNET | -0.0224 | -0.0040 | -0.0186 | -0.0202 | 0.0416 | 0.0600 | 0.0701 | 0.0612 | 0.0517 |
| ESSENTNET × AMAZ | 0.1651** | 0.2160*** | 0.1946*** | 0.2065*** | 0.1397* | 0.1829 | 0.1752 | 0.1856 | 0.2156* |
| CODESHARE |  |  |  |  |  | -0.0932 | -0.0851 | -0.1099 | -0.0952 | -0.0868 |
| COVID | -0.0201 | -0.2845*** | -0.2350*** | -0.2588*** | -0.2568*** | -0.3385*** | -0.4770*** | -0.3612*** | -0.5490*** |
| COVID × DIST |  | 0.0328*** | 0.0173*** | 0.0199*** | 0.0190*** | 0.0287*** | 0.0370*** | 0.0301*** | 0.0432*** |
| COVID × EMERGAID |  |  | 0.0268*** | 0.0320*** | 0.0324*** | 0.0368*** | 0.0600*** | 0.0382*** | 0.0692*** |
| COVID × SOCINCL |  |  |  | -0.0133** | -0.0119** | -0.0164** | -0.0224** | -0.0197*** | -0.0395*** |
| Estimator | IV-LASSO | IV-LASSO | IV-LASSO | IV-LASSO | IV-LASSO | IV-LASSO | IV-LASSO | IV-LASSO | IV-LASSO |
| Alternative Version | No | No | No | No | No | No | COVID | SOCINCL | SUBSET |
| City Clusters | 88 | 88 | 88 | 88 | 88 | 88 | 88 | 88 | 88 |
| Region Controls | 5/5 | 5/5 | 5/5 | 5/5 | 5/5 | 4/5 | 4/5 | 4/5 | 3/5 |
| Region/Interact Controls | No | No | No | No | No | 27/70 | 20/70 | 26/70 | 21/70 |
| IV Count | 8/163 | 11/115 | 17/163 | 17/163 | 17/163 | 9/163 | 7/163 | 9/163 | 7/163 |
| AIC Statistic | -551 | -633 | -666 | -667 | -664 | -629 | -611 | -612 | -519 |
| BIC Statistic | -473 | -550 | -578 | -574 | -566 | -411 | -425 | -398 | -334 |
| Adj R2 Statistic | 0.3842 | 0.4474 | 0.4713 | 0.4729 | 0.4712 | 0.4645 | 0.4464 | 0.4513 | 0.3759 |
| RMSE Statistic | 0.1649 | 0.1561 | 0.1526 | 0.1522 | 0.1524 | 0.1506 | 0.1539 | 0.1526 | 0.1634 |
| RMSE CV Statistic | 0.1594 | 0.1588 | 0.1599 | 0.1605 | 0.1595 | 0.1525 | 0.1560 | 0.1530 | 0.1529 |
| Nr Observations | 763 | 763 | 763 | 763 | 763 | 763 | 763 | 763 | 763 |

*Notes: Estimation results produced by the instrumental variables, post-double-selection LASSO-based methodology of Belloni et al. (2012, 2014a,b), with fixed effects (IV-LASSO). Post-LASSO estimation is performed with a Two-Stage Least Squares, fixed-effects, procedure with standard errors robust to heteroskedasticity and autocorrelation. LASSO penalty loadings account for the clustering of origin city. Control variables' estimates omitted. Blank cells indicate that the variable was not used. DIST, PVAR, INCPRE, DENSPRE, LEISPRE, ESSENTNET, ESSENTNET × AMAZ, CODESHARE, and COVID not penalized by LASSO. Endogenous variables: PVAR, COVID, and CODESHARE. In Column (9), SUBSET means that the model employs alternative versions of the following subset of variables: COVID, SOCINCL, INCPRE, and DENSPRE. A stratified bootstrapping procedure was performed (strata = city) to correct the standard errors to correct for the fact that SOCINCL is obtained from a prediction with previously estimated parameters. Cross-validation was performed with a 4-fold procedure. P-value representations: \*\*\*p<0.01, \*\* p<0.05, \* p<0.10.*

Also notable is the complete flip in the coefficients of the DIST dummy variables' signs (DIST 500–1000, DIST 1000–1500, ...): unlike Column (1), they are all negative in Column (2). This apparent contradiction does not alter the finding in Column (1) that PAXVAR is positively correlated with distance. With the interaction term COVID × DIST, the full effect of distance on PAXVAR in (2) is equal to the coefficient of the respective distance dummy *plus* the coefficient of the interaction term multiplied by COVID. This second term outweighs the first for the majority of the observations, such that the estimated full influence of distance on PAXVAR is still positive, consistent with the results in Column (1).[31]

In Columns (3), (4), and (5), we introduce regressors COVID × EMERGAID, COVID × SOCINCL, and CODESHARE, respectively. These variables have estimated coefficients that were positive, negative, and nonsignificant, respectively. From these specifications, we can infer that: 1) the emergency aid granted by the government in response to the pandemic had an attenuation effect on the relationship between COVID and PAXVAR; 2) the routes marked by greater social inclusion in the early 2010s had a more significant impact on demand during the first quarter of the COVID-19 pandemic crisis; and 3) the codeshare agreement between Latam and Azul airlines does not seem to have produced any immediate soothing effect on demand drop.

We consider the specification (5) as our preferred PVAR model. With this model, we estimate the full elasticity of PAXVAR with respect to COVID to be between -.0082 and -.0880, considering the 25 and 75 percentiles of DIST, EMERGAID, and SOCINCL. Furthermore, as we are particularly interested in the interpretation of the estimates of the COVID × SOCINCL variable, we also report its associated full effect. Based on our estimation results, we have evidence that ceteris paribus, a 1% increase in SOCINCL is associated with a 0.153% to 0.166% decrease in air travel demand in the sample period.[32] This result has important implications for airlines' corporate policy that may help their market repositioning and business model adjustments in the challenging years to come. With fewer passengers from the middle class, demand tends to shrink and become less price elastic. However, with the recent phenomenon of greater substitutability of air travel for virtual meetings by corporate passengers, the demand from this mainstream market segment of business travel may also diminish. Consequently, it becomes a matter of market survival for carriers to find ways to foster non-business passenger air travel demand. Our results may provide evidence in favor of the necessity of expanding the low-cost business model to recover at least part of the profitability lost in 2020 from the exploitation of leisure travel segments.

---

[31] See a discussion in Kennedy (2005, Example 8). Note that the number of observations with positive sign for the full effect of distance on PAXVAR increases in the more complete specifications of Table 2.

[32] These estimates are statistically significant at the 5% level. To compute the estimates, we calculate the marginal effect of SOCINCL on PAXVAR, extracted at the percentiles 25 and 75 of COVID.

*4.1 Robustness checks*

We consider some robustness checks to inspect the sensitivity of our main estimation results displayed in the fifth columns of Tables 2 and 3. We present the results of these experiments in Columns (6), (7), (8), and (9) of each table.

First, we check the robustness of the PAX model. In Column (6) of Table 2, we employ different metrics for some of the key variables of our PAX model. Unlike the remaining columns of Table 2, the model in (6) employs alternative versions of PBUS, TOUR, NET, LEIS, CELL, LOAN, and DEBT, as described in the Appendix. Note that it represents a relevant perturbation, with changes in more than half of the regressors. As a result of this experiment, the vast majority of variables kept their estimated coefficients relatively stable. The only change in the results concerns the LOAN variable, whose coefficient becomes positive and statistically significant. In the other robustness checks, this variable does not follow such a result, becoming nonsignificant in most specifications. We consider this particular estimate of LOAN an isolated finding, whose case should be further investigated in future studies.

In Columns (7)–(9), we introduce a broad set of controls to better approach the unobserved factors that may influence air travel demand during the pandemic. In Column (7), we introduce 48 local temperature controls;[33] in Column (8), we add 300 region-specific seasonal dummy variables;[34] and finally, in Column (9), we introduce 102 panel time controls. Note that in the latter estimation, due to perfect collinearity with the time dummies, the TREND and TREND × REC variables are not included in the regressor set. All the proposed controls were included in the LASSO penalization procedure to avoid overfitting. The final step of the estimation reported in Column (9) comprises 325 controls (11 local temperatures + 242 region/seasons + 72 times) selected by the LASSO estimation, from the 450 (48 + 300 + 102) originally assigned to the model. In all these robustness checks, the key results in Column (5) remain roughly unchanged.

Second, we check the robustness of the PAXVAR model. In Column (6) of Table 3, we insert a set of interaction terms as controls for the economic conditions of the local economies during the pandemic. To construct these terms, we multiply each regional dummy by the geometric mean and the simple mean of the gross domestic product and the population of the endpoints of each route. We then multiply each of these new variables by dummies of trunk routes (i.e., routes between state capitals) and dummies of regional routes (i.e., routes between country towns or between a capital and a country town). In total, we create 70 regions-interacted controls to capture the socioeconomic idiosyncrasies of each route's local economies in the sample period. Again, the proposed controls are included in LASSO's penalization for model selection. The procedure selected 27 controls from 70 subjects originally assigned. The results of this robustness check are roughly the same as those in Column (5).

---

[33] We collect local temperature data from the Meteorological Database of the Brazilian National Institute of Meteorology (INMET). These data allow building variables with a monthly variation. We create proxies for origin and destination minimum, mean, median, and maximum local temperature. We also create gravity versions of the proxies, namely, the minimum, maximum, and geometric mean between the origin and destination of each route at each period.

[34] This corresponds to 25 possible pair-of-regions times 12 months.

Finally, in Columns (7)–(9) we employ alternative versions of COVID, SOCINCL, INCPRE, and DENSPRE. The description of the alternative versions of the regressors is provided in the Appendix. Note that in these experiments, we also insert the region-interacted controls developed for Column (6). In all of the proposed experiments of Columns (7), (8), and (9), most of the estimated coefficients remain unchanged, thus providing further evidence of the robustness of our key findings.

## 5. Conclusion

### *5.1. Summary and results*

This study developed an empirical framework to pinpoint some of the drivers of the sharp decline in domestic air transport demand in Brazil since the coronavirus outbreak began. More specifically, we focused on the pandemic's impact on the routes linking cities marked by higher social inclusion. First-time consumers from the emergent classes have entered the air travel market with more intensity since the mid-nineties. The proposed methodology aimed to identify air transport markets possibly characterized by greater social inclusion and inspect their behavior during the first quarter of the pandemic period. Our contribution lies in assessing the possible impacts on the air travel demand of social inclusion indicators at the local economy level, such as income distribution, credit availability, and access to the Internet. Another contribution is the approach to address endogeneity regarding the possible COVID-19 contagion through air travel, thus avoiding inconsistent estimation of demand during the virus outbreak.

We produce evidence of the air travel market's differentiated reactions to the pandemic shock according to their social inclusion intensity. We find strong evidence that the demand plunge on routes marked by short haul and low density was the most dramatic. Moreover, we find suggestive evidence that business-oriented routes are impacted more strongly than leisure ones. We also estimate that a market with 1% higher social inclusion is associated with a roughly 0.153% to 0.166% more pronounced decline in demand during the pandemic. Our results show that the markets most benefited by greater social inclusion in Brazil are the most vulnerable to COVID-19 persistence. Finally, we find a positive correlation between COVID-19 cases and the error term of demand variation during the first quarters of the pandemic. We tackle this endogeneity by employing a LASSO-based IV approach.

*5.2. Policy implications*

Our findings provide some groundwork for taking stock of the adjustments in business models that airlines need to hasten to return to the pre-pandemic traffic levels. In particular, our evidence underpins the view that markets that are more reliant on social inclusion should be targeted with deeper discounts to bring back price-sensitive consumers. Carriers that react more quickly to these new market conditions are likely to get a head start in the post-pandemic environment. In addition, we conclude that the public policies implemented to relieve the problems caused by the pandemic to the air transportation sector had conflicting results. On the one hand, the emergency financial aid grant was successful to maintain some activity in the local economies in order to moderate the effects of COVID-19 on the demand for air transport; on the other hand, the launch of an "Essential Air Network" by the government and the major airlines apparently was not effective in mitigating the intensity of the fall in demand for air travel. Greater articulation of public and private policies could have enhanced the effectiveness of those measures.

*5.3. Limitations and possible extensions*

In this study, due to data availability, we worked with separate data samples to address the impact of the COVID-19 pandemic on air transportation in Brazil. The ideal approach would be to work with a single, combined dataset to exploit how the pandemic may have induced exogenous variation in the air travel demand change. We believe that future studies could address this limitation with a more complete dataset. Additionally, we emphasize that our analysis is restricted to the first two quarters of the pandemic period. We believe that the inclusion of data from other periods after those used in this sample could reveal other phenomena and serve as an empirical test of the evidence raised in our study. Finally, we consider that applying to realities of other emerging countries that have also recently experienced greater social inclusion could contribute to the findings of this work being assessed in a more general international context.

**Appendix A. Description of model variables**

Equation (1) 's variables:

- $PAX_{k,t}$ is the number of total air tickets sold by airlines on the city-pair (in logarithm). Source: ANAC's Airfares Microdata.[35]

- $INC_{k,t}$ is a proxy for mean income. It is equal to the per capita gross domestic product's geometric mean between the origin and destination cities (inflation-adjusted local currency values, in logarithm).[36] Source: IBGE.

- $P_{k,t}$ is the mean airline price on the city-pair (inflation-adjusted local currency values, in logarithm). Source: ANAC's Airfares Microdata.

- $PBUS_{k,t}$ is a proxy for the mean bus price on the city-pair (inflation-adjusted local currency values, in logarithm). This variable is centered on a reference of regulated bus yield - equal to 0.1524 BRL per kilometer in July 2015. Its variations across time - before and after July 2015 - are city-specific, proxied by the bus prices inflation per major city computed in Brazil's consumer price index. For the medium and small cities, we utilized the countrywide bus price inflation. For each route, we calculate the geometric mean yield between origin and destination cities and then multiply this mean by the respective distance in kilometers to create a proxy for mean bus price. Sources: ANTT and IBGE.

- $TOUR_{k,t}$ is a proxy for the size of the tourism segment on the route. It is equal to the number of total charter flights on the route divided by the maximum population between the origin and destination cities (in logarithm). This ratio is scaled in thousand flights per million population. Sources: ANAC's Air Transport Statistical Database[37] and IBGE.

---

[35] See details at www.nectar.ita.br/avstats/anac_airfares.html

[36] When defining a "city" to calculate socioeconomic indicators, we consider the full mesoregion to approach its airports' catchment area. During the sample period, IBGE employed the concept of "mesoregion" - a grouping of cities in the same region for statistical purposes. The concept has been revised and called "intermediary geographic regions" since 2017.

[37] See www.nectar.ita.br/avstats/anac_statdata.html.

- $NET_{k,t}$ is a proxy for the convenience and intrinsic quality of the flight network available to passengers from the endpoint cities (in logarithm).[38] This variable is equal to the geometric mean of the number of network points (cities) served with flights from/to the origin and destination cities. Source: ANAC's Air Transport Statistical Database.

- $P_{k,t} \times LEIS_k$ is an interaction variable, where $LEIS_k$ is a measure to capture the periods in which a higher proportion of leisure passengers travel on the city pair. With $P_{k,t} \times LEIS_k$, we aim to control from the varying price-elasticities of demand according to the intensity of leisure traffic on the route. We set this variable equal to a dummy of the high-season period - the Summer in the Southern hemisphere, from December to March.

- $TREND_t$ is a time trend variable to control for the national level (systemwide) demand evolution;

- $TREND_t \times REC_t$ is an interaction variable aimed to control a possible structural break in the national trend in demand due to Brazil's mid-2010s recession. $REC_t$ is a dummy variable assigned with one from April 2014 until December 2016 - a technical recession period.[39]

- $HDI_{k,t}$ is a proxy for social inclusion on the city-pair market. It is equal to the geometric mean of a proxy for the Human Development Index (HDI) between the origin and destination cities (in logarithm).[40] According to the United Nations Development Programme (2020, p. 244), the HDI is "(…) *a composite index measuring average achievement in three basic dimensions of human development—a long and healthy life, knowledge and a decent standard of living*." To compute this variable, we extract a monthly interpolation of the yearly available, city-specific, Firjan's Municipal Development Index (IFDM). According to Avelino, Bressan, and Cunha (2013), the IFDM is a simple mean of indexes assessing the three prominent United Nations' human development areas, with indexes for "Employment and Income" (decent standard of living), "Education" (knowledge), and "Health" (long and healthy life). All indexes have a theoretical range from 0 to 1 - the closer to 1, the higher the city's development levels. In practice, no index reaches neither of the extreme values. To compute our HDI measure, we aggregate Firjan's IFDM - which is city-specific - to the "mesoregion" (grouping of cities) level by extracting a weighted mean with the population's size as weights. If social inclusion is a driver of the air transport demand in Brazil during the sample period (2010-2018), HDI should be positively correlated with PAX.

---

[38] McCaughey and Behrens (2011) describe that the size of the flight network may be a measure of the quality of the airline frequent flier programs concerning the extent of miles earning and award possibilities.

[39] See "Brazil's economy slips into recession", August, 29, 2014, available on www.ft.con.

[40] In all our local economies' metrics, we employ the concept of mesoregion, i.e., a grouping of cities that belong to the routes' endpoint airports' catchment areas.

- HDI (DSL)$_{k,t}$ is a proxy for social inclusion with respect to the status of income equality on endpoints of the city pair. It is based on the Human Development Index (HDI) 's dimension for "decent standard of living" (DSL). This variable is equal to the geometric mean of the index extracted between the origin and destination cities (in logarithm). The procedure for the construction of this variable is the same as HDI$_{k,t}$. If social inclusion is a driver of air transportation demand, then ceteris paribus, HDI (DSL) should be positively correlated with PAX.

- CELL$_{k,t}$ is a proxy for digital inclusion. According to the United Kingdom's NHS, the definition of "digital inclusion" includes 1. being able to use digital devices, such as computers and smartphones, 2. access to the Internet, and 3. digital services designed to meet the users' needs.[41] In our case, CELL$_{k,t}$ is equal to the geometric mean of the number of cell phones per 100 inhabitants of the endpoint cities of a route (in logarithm). Source: ANATEL.

- DEBT$_{k,t}$ and LOAN$_{k,t}$ are proxies for financial inclusion. According to the World Bank, financial inclusion means "that individuals and businesses have access to useful and affordable financial products and services that meet their needs - transactions, payments, savings, credit and insurance - delivered in a responsible and sustainable way"[42], therefore being a relevant booster of economic prosperity. In our model, LOAN$_{k,t}$ is an indicator of households' access to bank credit. García-Escribano (2013) describes that total credit to GDP rose significantly in Brazil during the 2000s, from 25 to about 49 percent of GDP, with the consumer credit categories experiencing strong growth rates. LOAN$_{k,t}$ is a per capita flow variable equal to the difference between the number of new loans and those paid off in the last twelve months (inflation-adjusted local currency values, in logarithm). It is equal to the geometric mean between the origin and destination cities. To account for the effect of new bank credit grants, we only consider the variable's positive values, setting it to zero when negative.[43] This variable is denoted in constant, inflation-adjusted, BRL (local currency) values. Source: Central Bank of Brazil.

- DEBT$_{k,t}$ is a proxy for household indebtedness. It is equal to the mean household bank debt per capita. This metric is a stock variable and includes current bank financing and loans. We consider the geometric mean of the endpoint cities of origin and destination (in logarithm). This variable is denoted in constant, inflation-adjusted, BRL (local currency) values. Source: Central Bank of Brazil.

---

[41] "Definition of digital inclusion", last edited April 21, 2020, available on digital.nhs.uk.
[42] "Financial Inclusion," available at www.worldbank.org/en/topic/financialinclusion.
[43] To compute the logarithm of this variable, we added one to it.

Equation (2) 's variables:

- PAXVAR$_k$ is the change rate in total sold airline tickets on the city-pair since the pandemic outbreak of coronavirus disease, considering the second and third quarters of 2020 (2020Q2–Q3) against the same period in 2019 (2019Q2–Q3). Source: ANAC's Airfares Microdata.

- DIST 500 - 1000, DIST 1000 - 1500, DIST 1500 - 2000, DIST 2000 HIGH are dummy variables taking the value one if the distance of city-pair market is, respectively, between the 500–1000 km, 1000–1500 km, 1500–2000 km, and equal or higher than 2000 km, and 0 otherwise.

- PVAR$_k$ is the change rate in mean airline price since the pandemic outbreak, considering the second and third quarters of 2020 (2020Q2–Q3) against the same period in 2019 (2019Q2–Q3). Source: ANAC's Airfares Microdata.

- INCPRE$_k$ is a proxy for the mean income previous to the pandemic. It is equal to the geometric mean of the per capita gross domestic product between the origin and destination cities. This variable is expressed in inflation-adjusted BRL (local currency) values in logarithm.[44] To compute this variable, we consider the second and third quarters of 2018 - source: IBGE.[45]

- DENSPRE$_k$ is a proxy for the mean route density observed before the pandemic. It is equal to the number of the total sold airline tickets on the city-pair (in logarithm). To compute this variable, we consider the second and third quarters of 2019. Source: ANAC's Airfares Microdata.

- LEISPRE is a proxy for the proportion of leisure passengers traveling on the city-pair previous to the pandemic. Source: EPL - National Logistics & Planning Company's 2014 passenger survey.

- ESSENTNET is a dummy variable taking the value one if the city-pair is listed among the routes of the "Essential Air Network," outlined by the airlines and the government to keep minimum operations levels in response to COVID-19, and 0 otherwise.[46] Source: ANAC.

- ESSENTNET × AMAZ is an interaction variable in which AMAZ is a dummy variable taking the value one if at least one of the endpoint cities is from the Amazonas state, at least one is a country town, and both are located in the North of Brazil. Many towns at Amazonas are inaccessible by road, which makes the respective routes natural candidates for being essential network points from the public policy standpoint.

---

[44] We utilize the concept of "mesoregions" when defining a "city." See the discussion of INC$_{k,t}$.

[45] We employ 2018 because the 2019 figures were not available.

[46] See "Measures adopted by Brazil in response to COVID-19" by the Brazilian government at the Second Virtual Meeting of General Directors of Civil Aviation of South America Region on the response to COVID-19, Lima, Peru, May 22, 2020. Available at www.icao.int/SAM/Documents/2020-VM4-COVID19.

- CODESHARE is a dummy variable taking the value one if the city-pair is listed among the routes operated by the Latam Airlines - Azul Airlines codeshare agreement. Source: Azul Investor Relations.

- $COVID_k$ is the maximum number of confirmed Coronavirus infection cases between the endpoint cities, from February 25 to November 30, 2020 (in logarithm). Source: Brazil's Ministry of Health COVID-19 Portal (covid.saude.gov.br).

- $COVID_k \times DIST_k$ is an interaction variable in which $DIST_k$ is the geodesic distance between origin and destination cities in kilometers, calculated using the Vincenty formula and the respective latitude and longitude coordinates (in logarithm).

- $COVID_k \times EMERGAID_k$ is an interaction variable in which $EMERGAID_k$ is a proxy for the effects of the emergency financial aid grant to citizens on the local economies as a response to the COVID-19 pandemic. The emergency financial aid was a benefit granted by the federal government from April to December 2020 to individual workers, individual microentrepreneurs, self-employed and unemployed, and benefitted over 65 million people in up to nine monthly payments between approximately 30% and 60% of a minimum wage. The main objective of the benefit was to provide emergency protection during the pandemic. $EMERGAID_k$ is equal to the maximum number of emergency financial aid grants per population cases (in logarithm). Source: Brazilian government's Transparency Portal (www.portaltransparencia.gov.br).

- $COVID_k \times SOCINCL_k$ is an interaction variable in which $SOCINCL_k$ is our proxy for social inclusion. It is an indicator of the long-term social inclusion trend in altering the passenger profile on the route since the early 2010s. It is computed as a prediction from Equation (1), being equal to $\frac{1}{T}\sum_t \widehat{SI}_{k,t}/\widehat{PAX}_{k,t}$, where $\widehat{SI}_{k,t} = \hat{\beta}_9 HDI(DSL)_{k,t} + \hat{\beta}_{10} CELL_{k,t} + \hat{\beta}_{11} LOAN_{k,t} + \hat{\beta}_{12} DEBT_{k,t}$, with the hat operator indicating predicted values and T equal to the number of periods. To avoid contaminating this variable's effect with the recession period of 2014-2016, we consider only predicted values until 2013. With this variable, we aim to estimate the differentiated impact of the COVID-19 pandemic on Brazil's domestic routes according to social inclusion.

# Appendix B. Description of alternative version of model variables

**Table 4. Alternative version of regressors - PAX model (Table 1).**

| Regressor | Main model | Alternative Version |
|---|---|---|
| **PBUS** | This variable is centered on a reference of regulated bus yield - equal to 0.1524 BRL per kilometer in July 2015. | Instead of the *reference of bus yield*, we employ an inflation-adjusted *bus price index* (base 100) computed from Brazil's consumer price index data base. It is equal to the geometric mean yield centered at 100 on a route-specific basis, and without multiplying by distance. |
| **TOUR** | The number of total charter flights on the route divided by the maximum population between the origin and destination cities. | The number of total charter flights on the route divided by the maximum *number of flights* between the origin and destination cities. |
| **NET** | The geometric mean (between O and D) of the number of network points (cities) served with flights. | The maximum (between O and D) of the number of network points (cities) served with flights. |
| **LEIS** | A dummy of the high-season period - the Summer in the Southern hemisphere, from December to March. | A dummy of the high-season period - the Summer in the Southern hemisphere, from December to March -, interacted with the *proportion of leisure passengers* traveling on the city-pair (Source: EPL - National Logistics & Planning Company's 2014 passenger survey). |
| **CELL** | The geometric mean (between O and D) of the number of cell phones per 100 inhabitants. | The maximum (between O and D) of the number of cell phones per 100 inhabitants. |
| **LOAN** | The geometric mean (between O and D) of the difference between the amount of new loans and those paid off in the last twelve months. | The maximum (between O and D) of the difference between the amount of new loans and those paid off in the last twelve months. |
| **DEBT** | The geometric mean (between O and D) of the household bank debt per capita. | The maximum (between O and D) of the household bank debt per capita. |

**Table 5. Alternative version of regressors - PAVAR model (Table 2).**

| Regressor | Main model | Alternative Version |
|---|---|---|
| **COVID** | The maximum number of confirmed Coronavirus infection cases between the endpoint cities. | The maximum number of confirmed Coronavirus *deaths* between the endpoint cities. |
| **SOCINCL** | To avoid contaminating this variable's effect with the recession period of 2014q2-2016q4, we consider only predicted values until 2013. | The recession began in the mid-2014, then it is debatable if we can use either 2013 or 2014 as a cutoff year. Instead of considering *2013* as a cutoff, we consider *2014*. |
| **INCPRE** | The geometric mean (between O and D) of the per capita gross domestic product. | The maximum (between O and D) of the per capita gross domestic product. |
| **DENSPRE** | It is equal to the number of total sold airline tickets on the city-pair in the second and third quarters of 2019. | Instead of using *only 2019* figures, we alternatively consider the city-pair specific mean of *2017, 2018, and 2019* (second and third quarters). |

## Appendix C. Descriptive statistics

### Table 6 - Descriptive Statistics of model variables

| Eq. | Variable | Description | Level of measurement | Metric | N. Obs. | Original scale | | | |
|---|---|---|---|---|---|---|---|---|---|
| | | | | | | Mean | Std. Dev. | Minimum | Maximum |
| (1) | PAX | airline tickets sold | city-pair | count | 65,452 | 5,436.37 | 12,425.45 | 100 | 203,831 |
| | INC | GDP per capita | geom. mean (O, D cities) | BRL constant values | 65,452 | 2,945.54 | 938.56 | 1,004.52 | 7,460.04 |
| | P | airline price | city-pair | BRL deflated | 65,452 | 473.40 | 214.35 | 45.96 | 2,047.70 |
| | PBUS | bus price | geom. mean (O, D cities) | BRL deflated | 65,452 | 189.70 | 127.98 | 18.30 | 641.13 |
| | PBUS (alt) | bus index price | geom. mean (O, D cities) | index (mean = 100) | 65,452 | 101.36 | 4.47 | 78.20 | 119.77 |
| | TOUR | charter flights per population | city-pair; max (O, D cities) | ratio (K per milion) | 65,452 | 51.63 | 213.77 | 0.00 | 10,689.45 |
| | TOUR (alt) | charter flights per city flights | city-pair; max (O, D cities) | ratio (per milion) | 65,452 | 90.57 | 453.32 | 0.00 | 23,728.81 |
| | NET | served cities | geom. mean (O, D cities) | count | 65,452 | 19.25 | 9.68 | 2.00 | 63.94 |
| | NET (alt) | served cities | max (O, D cities) | count | 65,452 | 38.87 | 17.07 | 2.00 | 74.00 |
| | LEIS | high season | systemwide | dummy | 65,452 | 0.33 | 0.47 | 0 | 1 |
| | LEIS (alt) | leisure passengers during high season | city-pair; systemwide | fraction × dummy | 65,452 | 0.13 | 0.21 | 0.00 | 0.94 |
| | TREND | time trend | systemwide | discrete sequence | 65,452 | 49.35 | 29.06 | 1 | 102 |
| | REC | recession (Apr 2014 - Dec 2016) | systemwide | dummy | 65,452 | 0.32 | 0.47 | 0 | 1 |
| | HDI | human development index | geom. mean (O, D cities) | index × 100 | 48,965 | 73.22 | 6.48 | 48.99 | 85.06 |
| | HDI (DSL) | human development index (DSL) | geom. mean (O, D cities) | index × 100 | 48,965 | 67.61 | 7.45 | 40.33 | 80.80 |
| | CELL | cell phones per population | geom. mean (O, D cities) | ratio × 100 | 65,452 | 96.29 | 39.80 | 9.41 | 266.46 |
| | CELL (alt) | cell phones per population | maximum (O, D cities) | ratio × 100 | 65,452 | 134.68 | 50.89 | 22.01 | 481.59 |
| | LOAN | new loans per population | geom. mean (O, D cities) | ratio × 100 | 65,452 | 1,497.20 | 3,491.69 | 0.00 | 53,203.53 |
| | LOAN (alt) | new loans per population | maximum (O, D cities) | ratio × 100 | 65,452 | 6,066.53 | 11,464.21 | 0.00 | 60,066.26 |
| | DEBT | household bank debt per population | geom. mean (O, D cities) | ratio | 65,452 | 527.11 | 463.37 | 14.34 | 4,141.37 |
| | DEBT (alt) | household bank debt per population | maximum (O, D cities) | ratio | 65,452 | 1,472.61 | 1,616.83 | 16.69 | 6,336.12 |
| (2) | PAXVAR | change in tickets from 2019 to 2020 | city-pair | change rate | 765 | -0.67 | 0.21 | -1.00 | -0.01 |
| | DIST 500 - 1000 | distance between 500 and 1000 km | city-pair | dummy | 765 | 0.30 | 0.46 | 0 | 1 |
| | DIST 1000 - 1500 | distance between 1000 and 1500 km | city-pair | dummy | 765 | 0.19 | 0.39 | 0 | 1 |
| | DIST 1500 - 2000 | distance between 1500 and 2000 km | city-pair | dummy | 765 | 0.14 | 0.35 | 0 | 1 |
| | DIST 2000 HIGH | distance higher than 2000 km | city-pair | dummy | 765 | 0.15 | 0.36 | 0 | 1 |
| | PVAR | change in price from 2019 to 2020 | city-pair | change rate | 765 | -0.27 | 0.25 | -0.80 | 2.09 |
| | INCPRE | GDP per capita previous to the pandemic | geom. mean (O, D cities) | BRL deflated | 765 | 36,205.37 | 11,060.59 | 15,152.16 | 72,434.62 |
| | INCPRE (alt) | GDP per capita previous to the pandemic | maximum (O, D cities) | BRL deflated | 765 | 47,477.85 | 18,987.11 | 15,775.61 | 91,453.09 |
| | DENSPRE | tickets previous to the pandemic | city-pair | count | 765 | 22,330.78 | 51,096.69 | 117.00 | 558,086.00 |
| | DENSPRE (alt) | tickets previous to the pandemic (2017-2019) | city-pair | mean over 3 years | 765 | 23,051.77 | 53,444.26 | 85.00 | 637,286.31 |
| | LEISPRE | leisure passengers previous to the pandemic | city-pair | fraction | 765 | 0.42 | 0.16 | 0.05 | 0.94 |
| | ESSENTNET | essential air network routes | city-pair | dummy | 765 | 0.08 | 0.27 | 0 | 1 |
| | AMAZ | amazon state routes | city-pair | dummy | 765 | 0.02 | 0.13 | 0 | 1 |
| | CODESHARE | codesharing between Latam and Azul | city-pair | dummy | 765 | 0.13 | 0.33 | 0 | 1 |
| | COVID | confirmed covid infections | maximum (O, D cities) | count | 765 | 268,393.97 | 290,974.48 | 25,505.00 | 931,341.00 |
| | COVID (alt) | confirmed covid deaths | maximum (O, D cities) | count | 765 | 10,012.13 | 11,530.76 | 576.00 | 34,408.00 |
| | DIST | distance | city-pair | km | 765 | 1,157.43 | 739.27 | 103.87 | 3,290.29 |
| | EMERGAID | emergency financial aid grants per population | maximum (O, D cities) | ratio × 100 | 765 | 25.97 | 4.74 | 17.00 | 35.48 |
| | SOCINCL | predicted social inclusion proxy (ref. 2013) | city-pair | fraction | 765 | 0.48 | 0.16 | 0.14 | 0.91 |
| | SOCINCL (alt) | predicted social inclusion proxy (ref. 2014) | city-pair | fraction | 765 | 0.47 | 0.16 | 0.14 | 0.92 |